\documentclass[12pt]{iopart}
\begin{document}
\title[Uniform Approximation for the Fidelity]
{A Uniform Approximation for the Fidelity in Chaotic Systems}
\author{Nicholas R. Cerruti and Steven Tomsovic}
\address{Department of Physics, Washington State University, Pullman, WA
99164-2814}
\date{\today}

\begin{abstract}
In quantum/wave systems with chaotic classical analogs, wavefunctions
evolve in highly complex, yet deterministic ways.  A slight perturbation
of the system, though, will cause the evolution to diverge from its
original behavior increasingly with time.  This divergence can be
measured by the fidelity, which is defined as the squared overlap of the
two time evolved states.  For chaotic systems, two main decay regimes of
either Gaussian or exponential behavior have been identified depending on
the strength of the perturbation.  For perturbation strengths
intermediate between the two regimes, the fidelity displays both forms of
decay.  By applying a complementary combination of random matrix and
semiclassical theory, a uniform approximation can be derived that covers
the full range of perturbation strengths.  The time dependence is
entirely fixed by the density of states and the so-called transition
parameter, which can be related to the phase space volume of the system
and the classical action diffusion constant, respectively.  The accuracy
of the approximations are illustrated with the standard map.
\end{abstract}


\maketitle

\section{Introduction}

In our contribution to this special issue on random matrix theory (RMT),
we address the subject of wave field evolution governed by linear wave
equations.  In particular, our interest is in the  sensitivity of
evolution to perturbations.  The treatment of this problem illustrates
two well-known, important lessons arising in RMT studies.  First, RMT and
semiclassical theories, when used in tandem, give a much stronger
theoretical approach to a problem than either was capable of on its
own~\cite{btu}.  Second, even though RMT is inherently unphysical in
certain ways, nevertheless, careful application leads to physically
meaningful predictions~\cite{brody}.  For example, using RMT to mock up
the evolution of a wave packet generates a dispersal into the entire
Hilbert space simultaneously.  Thus, it lacks a dynamical continuity
property that a semiclassical theory would easily reveal as absurd.  It
turns out that for a wave mechanical system with a chaotic underlying ray
dynamics, there is a logarithmic time scale dependence on the wave
vector~\cite{zas,bb,bbtv}, and RMT does not ``know'' about this; i.e.~its
results cannot be trusted for wave propagation up to this time scale.
Yet, beyond the so-called logtime (where the semiclassical dynamics
increases in complexity exponentially~\cite{th}), RMT can be a
powerful tool for understanding the statistical nature of the dynamics.
In what follows, we take advantage of the RMT-semiclassical
complementarity and avoid making inferences about pre-logtime dynamics.

The study of the sensitivity of wave field evolution to perturbation has
recently attracted a great deal of
interest~\cite{jal:01,jac:01,cer:02,pro:02,cuc:1,cuc:2,ben:02} for several
fundamental reasons, i.e.~decoherence, reversibility, tomography...  In
the case in which the basic system is weakly coupled to an environment,
the sensitivity relates to the manner in which the system
decoheres~\cite{zurek}.  The recent focus on quantum computation has
greatly increased the attention given to this issue~\cite{nielsen:00}.
Another motivation has been the reversibility of wave field evolutions
such as the spin polarization echoes in nuclear magnetic
resonance~\cite{pastawski:95}.  For a broad class of problems,
propagating forward in time with two slightly different systems is
mathematically equivalent to propagating forward with one system and
reversing the dynamics with another for the same propagation time.
Another application is to problems of waves propagating through random
media (WPRM), the medium is likely partially characterized and often
slowly evolving; twinkling of starlight passing through the atmosphere is
an example.  The interest is often in performing tomography or at least
knowing what information can be deduced about the
medium~\cite{flatte:79}.  Similar statements can be made regarding the
study of disordered systems~\cite{altshuler:85}.  Lastly, we mention that
problems involving simple, chaotic systems or even strongly interacting,
many-body systems usually require taking measurements as a function of
some system parameter.  For these systems, single measurements may almost
seem like random number generation, whereas it turns out that important
information can be extracted from the parameter variations of the
measurements~\cite{meso}.  The trick is to know what information therein
exists.  Our focus here will remain on the wave field evolution behavior
without getting into specific physical applications.

Introduced nearly twenty years ago, a natural measure of the sensitivity
of wave field evolution is the fidelity, which is defined as the squared
overlap of some initial state propagated forward in time with two slightly
different systems~\cite{peres:84}.  For normalized wavefunctions, it
begins at unity and typically decays to some small value.  In the spin
polarization experiments previously cited, it was observed that a change
in the functional form of the fidelity's decay occurred depending upon
the dipolar interactions~\cite{usaj:98}.  The decay exhibited either
Gaussian or exponential behaviors and in some instances both.  This
mirrors known behavior found in simple, quantized chaotic
systems~\cite{jac:01,cer:02}; note that there also exists some work on
integrable, near-integrable, and mixed phase space
systems~\cite{prosen:02,jacquod:02,cerruti:03} which exhibit other
behaviors.  For extremely weak perturbation strengths, quantum
perturbation theory along with some simplifying assumptions quickly lead
to the prediction of a Gaussian decay~\cite{peres:84}.  For larger
perturbation strengths, this approach fails, and a Fermi Golden Rule
regime emerges which implies an exponential decay behavior~\cite{old}.
Semiclassical theory gives this decay rate in terms of a classical action
diffusion constant~\cite{cer:02}.  If the perturbation strength is even
larger, at some point the fidelity decay rate saturates because it is
impossible to decay faster than the logtime~\cite{jal:01}; this is known
as the Lyapunov regime because the rate is independent of the perturbation
strength, and often determined by the Lyapunov exponent.

Regardless of the strength of the perturbation or the nature
of the dynamics, for very short times the fidelity is quadratic.  This
regime which can be derived by using time dependent perturbation theory
may be extremely short in time and  difficult to observe.  Likewise, in
the long time limit, the fidelity will saturate to the inverse of the
effective number of states in the system.  Both of these effects decrease
in importance in the asymptotic limit, and are disregarded in what
follows; the Lyapunov regime is also disregarded since that involves
pre-logtime dynamics.

In this article we concentrate on developing a uniform approximation
that encompasses both the Gaussian and exponential decays as well as the
intermediate regime for simple chaotic systems; and we restrict our
attention to the Schr\"odinger wave equation.  The fidelity undergoes a
crossover from exponential to Gaussian behavior as a function of time,
and the value of the perturbation strength determines whether the
crossover is visible.  In fact, it turns out that the only information
residing in the decay of the fidelity is the density of states and the
transition parameter of RMT~\cite{fkpt}; the spreading width could also
substitute for the transition parameter.  Alternatively, using
semiclassical arguments, one could also say that the only information
contained is the phase space volume and the classical action diffusion
constant~\cite{cer:02}.

The paper is organized as follows.  In the next section we review
the Gaussian and exponential Fermi Golden Rule regimes found in
simple quantized chaotic systems.  In Sect.~3, a uniform approximation is
derived using RMT and a degenerate perturbation theory.  The theory is
applied to the quantized standard map in Sect.~4.  It is found to give an
excellent approximation of the quantum results.  We then conclude with
some discussion and closing remarks.

\section{The Fidelity}

Consider an evolving quantum wave function $| \alpha_\lambda(t) \rangle$
where $\lambda$ defines the parameters of a Hamiltonian, $H(\lambda)$.
The overlap of the same initial state, $| \alpha \rangle$,
propagated via two different Hamiltonians specified by $\lambda_1$ and
$\lambda_2$ is
\begin{equation}
      A_\lambda(\epsilon; t) = \langle \alpha_{\lambda_1}(t)
      | \alpha_{\lambda_2}(t) \rangle
      = \langle \alpha | \hat{U}^\dag_{\lambda_1}(t)
      \hat{U}_{\lambda_2}(t) | \alpha \rangle
\end{equation}
where $\hat{U}_{\lambda}(t)$ is the unitary evolution operator, $\epsilon
=\lambda_1 - \lambda_2$, and $\lambda = (\lambda_2 + \lambda_1) / 2$.  In
the limit that the system is strongly chaotic, the statistical properties
of the overlap are independent of $\lambda$, so we omit the $\lambda$
subscript from here on.  The fidelity is just the absolute square overlap
\begin{equation}
      {\cal C}(\epsilon; t) = |A(\epsilon; t)|^2
\end{equation}
It is sufficient to work with the overlap itself in developing the theory
since the operation of squaring only introduces cross-correlation effects
higher order in $\epsilon$ that can be ignored.

By inserting complete sets of states, the overlap can be rewritten in terms
of the eigenenergies and eigenstates of the perturbed and unperturbed
systems as
\begin{eqnarray}
      A(\epsilon; t) &=& \sum_n \sum_m \langle \alpha(0) |
        n_{\lambda_1} \rangle \langle n_{\lambda_1} |
        m_{\lambda_2} \rangle
        \langle m_{\lambda_2} | \alpha(0) \rangle \nonumber \\
      && \times \exp\{-\rmi[E_m(\lambda_2) - E_n(\lambda_1)] t / \hbar\}
\end{eqnarray}
The Bohigas-Giannoni-Schmit conjecture~\cite{bgs} and later works
suggest that for a chaotic system, it is appropriate to apply RMT
arguments to understand its statistical behavior.  Here, as
mentioned already, one has to add the caveat - as long as $\epsilon$ is
weak enough that ${\cal C}(\epsilon; t)$ decays on a much longer time
scale than the logtime.  The first property we would like to invoke is
the invariance of the ensemble over changes in bases.  Thus, any rotation
to a different initial state must generate the same decay behavior for the
fidelity.  It suffices to make an average over a complete set of initial
wave functions $| \alpha(0) \rangle$ to obtain
\begin{eqnarray}
\label{aeps}
      \overline{A(\epsilon; t)}
        &=& {1 \over N} \sum_n \sum_m
        |\langle n_{\lambda_1} | m_{\lambda_2} \rangle|^2
        \exp\{-\rmi[E_m(\lambda_2) - E_n(\lambda_1)] t / \hbar\}
\end{eqnarray}
where $N$ is the number of states in the complete set.  RMT has been
shown to be a strongly ergodic theory~\cite{brody,pandey}, which in this
case implies that the fluctuations of the quantity $\delta A =
A(\epsilon; t) - \overline{A(\epsilon; t)}$ vanish with increasing matrix
dimensionality as $N^{-1}$.

The averaged overlap may be decomposed into a diagonal and off-diagonal
part
\begin{eqnarray}
\label{aeps1}
      \overline{A(\epsilon; t)}
        &=& {1 \over N} \sum_n
        |\langle n_{\lambda_1} | n_{\lambda_2} \rangle|^2
        \exp[-\rmi \Delta E_n  t / \hbar] \nonumber \\
      &+& {1 \over N}
        \sum_n \sum_{m \ne n} |\langle n_{\lambda_1} |
        m_{\lambda_2} \rangle|^2 \exp\{-\rmi[E_m(\lambda_2)
        - E_n(\lambda_1)] t / \hbar\}
\end{eqnarray}
where $\Delta E_n = E_n(\lambda_2) - E_n(\lambda_1)$ is the change in an
eigenenergy due to the perturbation.  For $\epsilon=0$, the off-diagonal
term exactly vanishes, the diagonal term equals unity, and the fidelity
does not decay as must be case.

\subsection*{The quantum perturbative regime}

The diagonal term is defined by adiabatically following an eigenlevel as
the system's parameters are continuously varied.  It may be further
evaluated by ensemble averaging
\begin{eqnarray}
     \left\langle \overline{A_{diag}(\epsilon; t)}\right\rangle
        &=& {1\over N} \sum_n\left\langle|\langle n_{\lambda_1} |
n_{\lambda_2} \rangle|^2 \exp\left[{-\rmi \Delta E_n  t \over
\hbar}\right] \right\rangle
\end{eqnarray}
It is most significant for extremely small perturbation strengths.  From
quantum perturbation theory, the lowest order shift of an eigenvalue
depends on the diagonal element of the perturbation whereas the rotation
of the eigenstates depends on off-diagonal elements and energy
differences.  Within RMT the perturbation matrix elements are
uncorrelated and thus to lowest order in $\epsilon$, the ensemble
averaging of the amplitudes and phases can be separated.  The average
over the phases is over a Gaussian probability density for the diagonal
matrix elements.  Also, the ensemble average of the amplitude is
independent of $n$.  Thus,
\begin{eqnarray}
\label{adiag}
       \left\langle \overline{A_{diag}(\epsilon; t)}\right\rangle
        &\approx& \left\langle|\langle n_{\lambda_1} | n_{\lambda_2}
\rangle|^2\right\rangle
        \exp(-\epsilon^2 \sigma^2_v t^2 / 2 \hbar^2)
\end{eqnarray}
where $\sigma^2_v$ is the variance of the diagonal matrix elements of
the perturbation $V$ defined by writing the Hamiltonian
locally as $H(\lambda)=H_0+\epsilon V$.  In the limit of a
differential $\epsilon$,
$\sigma^2_v$ is equal to the variance of the level slopes by the
Hellmann-Feynman theorem~\cite{feyn}.  If the perturbations are
sufficiently small such that the eigenfunctions do not  significantly
vary, i.e. $\langle n_{\lambda_1} | m_{\lambda_2}
\rangle
\approx
\delta_{nm}$, then the off-diagonal terms will be negligible and may be
ignored.  Hence, the fidelity has a Gaussian decay in time in the weak
perturbation regime.

The diagonal matrix element (level slope) variance is not a free
parameter, and can be given an elegant interpretation through a
semiclassical approach~\cite{lct,cllt,keating}.  It is given by
\begin{equation}
      \label{diag2}
      \sigma^2_v \approx {2 g K(E) \over \pi \hbar \overline{d} \beta}
\end{equation}
where $2g / \beta$ is the number of classical orbits with identical action
and $\overline{d}$ is the total mean density of states. The
index $\beta = 1$ for time-reversal-invariant systems and
$\beta = 2$ for  time-reversal-breaking systems.  $K(E)$ is the classical
action diffusion constant on the energy surface $E$ given by~\cite{bgos}
\begin{equation}
      \label{k(e)}
      K(E) = \int_0^\infty \left\langle V({\bf p}(0), {\bf q}(0); \lambda)
      V({\bf p}(t), {\bf q}(t); \lambda)  \right\rangle_{po} dt
\end{equation}
In this expression the averaging is defined over the primitive periodic
orbits of long period.  The physical picture is that the action difference
of two long orbits continuously deformable into each other as a function
of $\epsilon$ is the result of a diffusive process.  As the orbit explores
the phase space ``randomly'', sometimes its action increases relative
to the other orbit, sometimes it decreases.  The variance of that
diffusion is proportional to the variance of the level slopes.

\subsection*{Semiclassical analysis}

The above analysis describes a very restricted range of perturbation
$\epsilon$ centered at zero.  A semiclassical approach without the
quantum perturbation limitations can be developed that is valid over
a much broader range.  The semiclassical construction of an evolving
wave function begins with the propagator
\begin{eqnarray}
\langle {\bf q} | \hat{U} | {\bf q}^\prime \rangle &\approx&
        \left( {1 \over 2 \pi \rmi \hbar} \right)^{d/2}
        \sum_j \left| \det \left( {\partial^2 W_j({\bf q}, {\bf q}^\prime; t)
        \over \partial {\bf q} \partial {\bf q}^\prime} \right)\right|^{1/2}
     \nonumber \\ && \qquad\qquad\qquad  \times \exp \left( \rmi W_j({\bf q},
{\bf q}^\prime; t) / \hbar - {\rmi \pi \nu_j \over 2} \right)
\end{eqnarray}
The phase is specified by the time integral of the Lagrangian
$W_j({\bf q}, {\bf q}^\prime; t)$ and and an index based on the properties
of the conjugate points (like focal points), $\nu_j$.

The overlap decay in terms of the propagator is
\begin{eqnarray}
      A(\epsilon; t) &=& \int \rmd{\bf q} \rmd{\bf q}^\prime
        \rmd{\bf q}^{\prime \prime}
        \langle \alpha | {\bf q} \rangle
        \langle {\bf q} | \hat{U}^\dag_{\lambda_1}|
        {\bf q}^\prime \rangle
        \langle {\bf q}^\prime | \hat{U}_{\lambda_2}|
        {\bf q}^{\prime\prime} \rangle \langle {\bf q}^{\prime\prime} |
        \alpha \rangle
\end{eqnarray}
For the reasons given in the discussion surrounding
equation~(\ref{aeps}), there can be no overlap decay dependence on the
specific initial state for chaotic systems as long as the decay is on a
time scale which is much longer than the logtime.  Since we have
explicitly excluded the Lyapunov regime from consideration in this paper,
we are free to choose any form for the initial state.  For convenience,
we take Gaussian wave packets
\begin{equation}
      \langle {\bf q} | \alpha \rangle = (\pi \sigma^2)^{-d/4}
      \exp \left[ -{({\bf q} - {\bf q}_\alpha)^2 \over 2 \sigma^2}
      + {\rmi {\bf p}_\alpha \over \hbar} ({\bf q} - {\bf q}_\alpha) \right]
\end{equation}
The initial coordinate of the actions are expanded about the
center of the wave packet
\begin{eqnarray}
      W_j({\bf q}, {\bf q}^\prime; t) = W_j({\bf q}, {\bf q}_\alpha; t)
      &-& {\bf p}_\alpha \cdot  ({\bf q}^\prime - {\bf q}_\alpha) +
\nonumber \\
&&{1 \over 2} ({\bf q}^\prime - {\bf q}_\alpha)^T \left.{\partial W_j({\bf
q}, {\bf q}^\prime; t) \over \partial {\bf q}^\prime \partial {\bf
q}^\prime }\right|_{{\bf q}_\alpha} ({\bf q}^\prime - {\bf q}_\alpha)
\end{eqnarray}
Changes in the amplitudes are ignored, since the most important
contribution to the overlap decay is each orbit's action change due to
its division by $\hbar$.  Applying stationary phase on the ${\bf q}$ and
${\bf q}^{\prime\prime}$ integrals, we obtain
\begin{equation}
      \label{eq:semi}
      A(\epsilon; t) \approx \int \sum_j c_j \exp \left[ {\rmi \over \hbar}
\Delta W_j({\bf q}, {\bf q}_\alpha; t) \right] \rmd {\bf q}
\end{equation}
where $c_j$ is a magnitude which disappears from the expressions ahead,
but could be deduced using the equations in the appendix of~\cite{cllt}.
The $\Delta W_j({\bf q}, {\bf q}_\alpha; t)$ are the action differences of
two rays that begin at ${\bf q}_\alpha$ and end at position ${\bf q}$ in a
time $t$ and are continuously deformable into each other; bifurcations
are neglected.  First order classical perturbation theory gives
\begin{equation}
      \Delta W_j({\bf q}, {\bf q}_\alpha; t) = \epsilon \int^t_0
      {\partial L_j({\bf q}, {\bf q}_\alpha; t^\prime) \over \partial \lambda}
      dt^\prime
\end{equation}
The stability matrix elements and the Lagrangian are evaluated along the
orbits of $H(\lambda)$.  In general, stationary phase integration cannot
be performed on the last integral because the action differences are less
than $\hbar$.

Again a statistical argument can be employed.  The weighted phases will be
nearly Gaussian distributed for the reason mentioned earlier that
long orbit action changes are part of a diffusion process~\cite{cer:02}.
The last integral and summation in equation~(\ref{eq:semi}) can be replaced by
a Gaussian integral.  Thus, we obtain
\begin{equation}
      A(\epsilon; t) \approx \exp(-\sigma^2_W / 2 \hbar^2)
\end{equation}
where $\sigma^2_W$ is the variance of the action differences and is given
by~\cite{lct,cllt}
\begin{equation}
\label{sw}
      \sigma^2_W = 2 \epsilon^2 K(E) t = (2 \pi \overline{d} \sigma^2_v)
      \epsilon^2 \hbar \beta t / 2 g
\end{equation}
with the same action diffusion constant as before.  In contrast to the
quantum perturbative argument giving a quadratic $t$-dependence in the
exponential, here the argument is linear.  The semiclassical argument
does not contain the action correlations necessary for the theory to
contain the quantization of the spectrum.  It therefore misses the
Gaussian contribution from the diagonal terms, and essentially evaluates
the off-diagonal contribution in the regime of larger perturbations as
though there is a continuum.  The last form given in the equation exhibits
the Fermi Golden Rule form expected for such circumstances.  Note that
the symmetry factors do not enter $K(E)$ (the symmetry dependences cancel
in the final form) in contrast to the diagonal term.

\section{A Uniform Approximation}

A straightforward argument determines the value of $\epsilon$ which
separates the exponential and Gaussian regimes.  We emphasize that this
is {\it not} the cross-over to exponential dependence beginning from the
quadratic time dependence derived using time-dependent perturbation
theory, but rather the relative strength of contributions from diagonal
versus off-diagonal contributions in equation~(\ref{aeps1}).  Note that
either dependence dominates over the time range in which the other's
argument of the exponential takes on the lesser value.   A cross-over
between the two regimes occurs when their arguments are equal.  Simple
algebra gives the this time scale $t^*=h\bar d \beta / 2g$, which is the
Heisenberg time, $\tau_H$, to within the symmetry factor.  Exponential
decay dominates if the decay is completed by $\tau_H$, and Gaussian decay
dominates if little decay has occurred by $\tau_H$.  In between it turns
out that the decay has components of both behaviors.  In terms of the
parameters, the question is whether
\begin{equation}
     \label{criter}
     \epsilon^2 {{? \atop >} \atop {< \atop ~}} {\hbar^2\over K(E)
     t^*}= {\hbar g\over \pi \bar d \beta K(E)}
\end{equation}
It turns out using the results derived next that near the equality, the
magnitude of the off-diagonal term is comparable to the diagonal term.

Returning to equation~(\ref{adiag}), it is an oversimplification to take the
amplitude equal to unity.  Instead, we define the fraction
\begin{equation}
     f = \left\langle |\langle n_{\lambda_1}|n_{\lambda_2}\rangle|^2
     \right \rangle
\end{equation}
and ahead give a theory for its value in the neighborhood of unity (small
$\epsilon$) using a degenerate perturbation theory and RMT.  It is worth
noting the broader significance of the quantity $f$.  It and some closely
related measures have previously been studied by several authors.
In~\cite{bruus:96}, the authors give the leading analytic correction from
unity for the case of broken time reversal symmetry, and infer the
asymptotic behavior for large $\epsilon$, and
in~\cite{alhassid:95,kusnezov:96}, a rough guess is given for the full
range of behavior consistent with numerical simulations with random
matrices.  These measures' interpretations as parametric correlators of
eigenstates, eigenstate components, or widths were emphasized.
Motivations for their introduction come from a broad variety of
possibilities of parametric dependence on controllable external or
uncontrollable variables such as electromagnetic fields, thermodynamics
quantities, or shape/geometric properties.  Physical realizations from
nuclear fission to quantum dots to atomic and molecular spectroscopy are
possible.  It is interesting to note that the same quantity is entering
into the theory of the fidelity, but only for transition values of
$\epsilon$ between the two domains giving exponential and Gaussian
decay.

The argument giving a Gaussian behavior for the phase average remains
valid to a regime where $f$ is quite small and the diagonal term
negligible.  It suffices to evaluate $f$ to complete a more broadly
applicable expression for the diagonal term.

The off-diagonal term is more subtle.  By normalization, the prefactor
must be equal to $1-f$, but the argument of the exponential needs to
be determined.  Recall that semiclassical dynamics is valid far beyond
the logtime, but breaks down on algebraic scales before
$\tau_H$~\cite{th,sth}.  In particular, this implies that in the limit of
a very weak perturbation, the exponential scale as previously derived
places the exponential behavior beyond $\tau_H$.  The semiclassical
expression turns out to require modification.  Since it is valid on
time scales shorter than $\tau_H$, we can expand the exponential to
the linear time term.  This implies that in this regime the argument of
the exponential multiplied by $1-f$ must be equal to the argument found
in the previous subsection.  Therefore, to a good approximation, the
overlap can be written
\begin{eqnarray}
      \label{eq:full}
      \overline{A(\epsilon; t)} &\approx& f \exp{-\epsilon^2 \sigma^2_v t^2
\over 2 \hbar^2}  + (1 - f) \exp{-\epsilon^2\pi \beta \sigma^2_v
\overline{d} t
\over 2\hbar g (1 - f) } \nonumber \\
&\approx& f \exp{-\epsilon^2 g K(E) t^2 \over \pi \beta \overline{d}
\hbar^3}  + (1 - f) \exp{-\epsilon^2 K(E) t \over \hbar^2 (1 - f) }
\end{eqnarray}
where the first form emphasizes the quantum information (matrix element
variance or level slope variance), and the second form emphasizes the
classical information (classical action diffusion constant).  The
transition between the two regimes, Gaussian and exponential, as the
strength of the perturbation is increased, is given by the fraction
$f$ remaining in the diagonal overlap factor.  The exponential decay rate
has been modified by the introduction of $1 - f$ in the denominator which
is necessary to offset the reduced normalization.  In the limit of large
perturbations where $f\rightarrow 0$ we recover the previous results for
the exponential Fermi Golden Rule decay, and in the small perturbation
limit, $f\rightarrow 1$, the Gaussian decay.

\subsection*{Degenerate perturbation theory within RMT}

To derive the leading corrections to $f$ for both time-reversal-invariant
and noninvariant cases, the most direct approach would be to use the
second order perturbation theory expression
\begin{equation}
     |\langle n_{\lambda_1} | n_{\lambda_2} \rangle|^2
      \approx 1 - \epsilon^2 \sum_{m \ne n} {|V_{mn}|^2
      \over [E_n(\lambda_1) - E_m(\lambda_1)]^2} + \cdots
\end{equation}
where $V_{mn} = \langle m_{\lambda_1} | V | n_{\lambda_1} \rangle$.
Within RMT, $V$ is a Gaussian random matrix with the correct
symmetries, Gaussian orthogonal ensemble (GOE), unitary ensemble (GUE),
etc...  We only consider the time-reversal-invariant and
time-reversal-noninvariant cases, GOE and GUE respectively.

Define an energy variable $s$ rescaled by the mean local spacing and a
unit variance variable $x$ for the perturbation matrix elements.  Then
the above sum can be reexpressed as the following double integral
\begin{equation}
\label{pert}
      f \approx
      1 - 2\Lambda \int^\infty_{-\infty} \int^\infty_0
      { x^2 \over s'^2} R_2(s') \rho(x;s') \rmd s' \rmd x
\end{equation}
where $\Lambda = \epsilon^2 v^2 \overline{d}^2_r$ is the transition
parameter and $v^2$ is the local variance of the off-diagonal matrix
elements of $V$~\cite{fkpt}.  $\overline{d}_r$ and $v^2$ are defined for
the irreducible representations of the system.  If there are symmetries,
then these quantity must be adjusted accordingly by dividing out these
symmetries,  eg.~$\overline{d}_r = \overline{d} / g$ and $v^2 = \beta
\sigma^2_v / 2$.  The function $R_2(s)$ is a sequence of
$\delta$-functions which determine the spacings, and is known as the
two-point density correlation function.  Similarly, the function
$\rho(x;s)$ is a sequence of $\delta$-functions whose intensities
determine the matrix elements.  Crudely speaking, taking the ensemble
average of the above expression is reduced to inserting the $R_2(s)$
function from the appropriate Gaussian ensemble, and taking $\rho(x;s)$
to be a zero-centered, unit variance, Gaussian probability density
independent of
$s$.

It was emphasized by French and co-workers that the transition parameter
$\Lambda$ was quite general~\cite{fkpt}.  Although it naturally emerged
from perturbation theories, its relevance extended to entire symmetry
breaking transitions and it is the only scale that can enter
those problems; a similar statement applies to parametric statistics as
well~\cite{as}.

It is straightforward to understand that for $s$ near zero, the integral
diverges for the GOE, and that this approach must fail to give proper
results for $f$.  The two-point correlation function is approximately
linear, $R_2(s) \approx \pi^2 s / 6$, so the above integral will diverge
logarithmically.  However, we can apply a degenerate perturbation theory
approach developed in~\cite{tomsovic:86,fkpt}; the degenerate theory
also gives an extra valid correction for the GUE.  By diagonalizing a two
by two matrix, the difference in eigenenergies is given
by~\cite{tomsovic:94}
\begin{eqnarray}
      \Delta E_n &=& {1 \over 2} \sum_m
        \{\sqrt{[E_n(\lambda_1) - E_m(\lambda_1)]^2 + 4 \epsilon^2 |V_{mn}|^2}
        - [E_n(\lambda_1) - E_m(\lambda_1)]\} \nonumber \\
      && \qquad \times \mbox{sign}(E_n(\lambda_1) - E_m(\lambda_1))
\end{eqnarray}
Using this result, equation~(\ref{pert}) can be reexpressed as
\begin{equation}
      f \approx
        1 - {1 \over \sqrt{2 \pi}} \int^\infty_{-\infty} \int^\infty_0
        \left(1 - {s' \over \sqrt{s'^2 + 4 \Lambda x^2}} \right)
        R_2(s') \exp(-x^2 / 2) \rmd s' \rmd x \nonumber \\
\end{equation}
Away from very small $s'$, the degenerate perturbation result is
unnecessary, but still gives the exact leading order results as a function
of $\Lambda$.  This integral is rather difficult to evaluate exactly, but
the correct leading order results can be extracted by splitting the
integral into two regions $[0,s]$ and $[s,\infty]$.  First, the
eigenstate overlap is written
\begin{eqnarray}
    1- f \approx
        && \left[{1 \over \sqrt{2 \pi}}
       \int^\infty_{-\infty}\int^s_0  \left(1 - {s' \over \sqrt{s'^2 + 4
       \Lambda x^2}} \right) R_2(s') \exp(-x^2 / 2) \rmd s' \rmd x \right.
       \nonumber \\
     && \qquad + \left.
       \int^\infty_s {2 \Lambda \over s'^2} R_2(s') \rmd s' \right]
\end{eqnarray}
where the power series for $R_2(s)$ is substituted in the first term, and
the exact function is left in the second term.  For our purposes, it is
sufficient to replace $R_2(s)$ in the first region by ${\pi^2 s/ 6}$,
which gives the GOE result, or by ${\pi^2 s^2 / 3}$, which gives the GUE
result.  The exact two-point correlation functions are~\cite{brody}
\begin{equation}
      R_2(r) =\left\{\matrix{ 1 - [s(r)]^2 + Js(r)Ds(r) & \qquad \mbox{GOE}
\cr   1 - [s(r)]^2 & \qquad \mbox{GUE} } \right.
\end{equation}
where $s(r) = \sin(\pi r) / \pi r$, $Ds(r) = \rmd s(r) / \rmd r$ and
$Js(r) = \int^r_0 s(t) \rmd t - 1/2$.  Next, the $[0,s]$ integrals are
evaluated exactly, and their asymptotic expansions kept up to (not
including) $O(\Lambda^2)$.  The other integral is also exactly integrated,
and its small-$s$ expansion is made.  When properly formulated and
the two evaluations summed, all the divergent terms less than
$O(\Lambda^2)$ vanish; also the
$s$-dependent terms coming from the $[0,s]$ integration vanish.  The
$s$-independent terms give the desired result.  The remaining
$s$-dependent terms arising from the $[s,\infty]$ integral vanish as
higher order terms in the power series of $R_2(s)$ are included in the
first integral.  In this way, the final result does not depend on the
value chosen for $s$. Carrying out this technique up to but not including
$O(\Lambda^2)$, we obtain
\begin{equation}
      \label{eq:beta}
      f \approx \left\{ \matrix{ 1 - {\pi^2 \Lambda \over 6}
     \left( 1 - \gamma - \ln \left| { \pi^2 \Lambda \over 4 }\right|
\right) & \qquad \mbox{GOE} \cr
       1 - {2 \pi^2 \Lambda \over 3} + {32 \pi \Lambda \over 9}
        \sqrt{2 \pi \Lambda} & \qquad \mbox{GUE} } \right.
\end{equation}
where $\gamma$ is the Euler constant.  The GOE results and the
correction term ($\Lambda^{3/2}$) for the GUE are new, and the leading
GUE term is consistent with reference~\cite{bruus:96}, which also gives
the asymptotic results
\begin{equation}
      \label{eq:beta2}
      f \sim \left\{ \matrix{ {1 \over \pi^2 \Lambda}
      & \qquad \mbox{GOE} \cr
       {1 \over \pi^2 \Lambda} & \qquad \mbox{GUE} } \right.
\end{equation}

The overlap ensemble average is expected to be universal and should
apply to all systems chaotic enough to expect RMT behavior.  Numerical
studies and heuristic arguments suggest that the overlap between
perturbed and unperturbed eigenstates is roughly a Lorentzian-like
function of the perturbation strength~\cite{alhassid:95,kusnezov:96}.  We
look for a dependence of the form
\begin{equation}
     \label{eq:lorentzian}
     f = {1 \over 1 + h(\Lambda)}
\end{equation}
Note that in the previously discussed forms, the expression for $f$ was
raised to the power $\beta$, $h(\Lambda) \approx a_\beta \Lambda$, and
approximate values of the constants $a_\beta$ were given by a numerical
fit.  Using perturbation theory Wilkinson and Walker~\cite{wilkinson:95}
gave an expression for the individual off-diagonal matrix elements which
has a compatible form.  Nevertheless, the $O(\Lambda\ln\Lambda)$ and
$O(\Lambda^{3/2})$ corrections plus the asymptotic expressions indicate
difficulties with their interpolation since it gets both limits of small
and large $\Lambda$ incorrectly.
Equations~(\ref{eq:beta}-\ref{eq:lorentzian}) are, of course,
insufficient to determine $h(\Lambda)$.  However, we suggest the
following as perhaps the simplest forms for somewhat improved
interpolations that are at least analytically consistent with the limiting
regimes of large and small $\Lambda$:
\begin{equation}
     \label{fl}
     h(\Lambda) \approx \left\{ \matrix{{\pi^2 \Lambda \over 6} \left( 1
     - \gamma - \left({1 + b_0 \pi^4\Lambda^2 \over 1 + b_1 \pi^2\Lambda +
b_0 \pi^4\Lambda^2} \right) \ln\left|{ \pi^2 \Lambda \over 4 +
\exp(5+\gamma)  \pi^2 \Lambda}\right| \right) & \qquad \mbox{GOE} \cr \cr
     {2\pi^2 \Lambda \over 3} \left({1 + {3 \over 2} c_0 \pi^2 \Lambda
\over 1 + {16 \over 3} \sqrt{{2\over \pi} \Lambda} + c_0 \pi^2 \Lambda }
\right)   & \qquad \mbox{GUE}}\right.
\end{equation}
The use of a quotient of two polynomials inserts a general function
with some fitting parameters that we determined with GOE and GUE Monte
Carlo simulations; see the appendix for the Monte Carlo details.  Our best
fits are: $b_0 = 0.932$, $b_1 = 0.618$, and $c_0 = 1.89$.
Figure~\ref{fig:interpolation} shows the differences between the Monte
Carlo results and the new interpolations along with the differences
between the Monte Carlo results and those from
references~\cite{alhassid:95,kusnezov:96}.  Even though the new
interpolations have the correct limiting forms, they still appear to have
some small, detectable deviations from the exact RMT predictions.
Perhaps, the main conclusion is that the previous interpolations do
surprisingly well given that they fail to get any of the limiting cases
correctly.

Again, semiclassical theory fixes the value of $\Lambda$ using the
arguments presented in the previous section.  Thus,
\begin{equation}
     \label{lam}
     \Lambda = {\epsilon^2 \beta \sigma^2_v \overline{d}^2 \over 2 g^2}
     = {\epsilon^2 \overline{d} K(E)  \over \pi  g \hbar}
\end{equation}
where the last form follows from equation~(\ref{diag2}).  In some simple
systems such as the standard map used in the next section, $K(E)$ can be
calculated analytically.  In fact, the criterium, equation~(\ref{criter}),
for  observing either Gaussian or exponential behavior can be written in
terms  of the transition parameter
\begin{equation}
     \label{criter_lambda}
      \pi^2 \Lambda {{? \atop >} \atop {< \atop ~}} {1 \over  \beta}
\end{equation}
It is remarkable that the same transition parameter that determines the
effects of weak symmetry breaking in RMT, gives precisely the
exponential-to-Gaussian cross-over for the fidelity.

Finally, the uniform approximation for the fidelity is the square of
equation~(\ref{eq:full}) with the results given in
equations~(\ref{eq:lorentzian}-\ref{lam}).  It could have turned out that
there were three parameters in the functional form of the decay, the
relative amount of Gaussian to exponential component ($f$), and the
scales in the arguments of the Gaussian and exponential functions.
Interestingly, other than the density of states (or phase space volume),
only one quantity determines all three potential parameters, that being
the classical action diffusion constant.  There is no other information
contained in the decay of the fidelity for a strongly chaotic system.

\section{Fidelity in the quantized standard map}

We use the standard map, which is a paradigm for chaotic systems, to
demonstrate our results for time-reversal-invariant systems.  For a
kicking strength $\lambda$ greater than 6 or so the  standard map is fully
chaotic.  The classical map is defined by
\begin{eqnarray}
      p_{i + 1} &=& p_i - (\lambda / 2 \pi) \sin (2 \pi q_i) \ \ \ \mbox{mod}(1)
        \nonumber \\
      q_{i + 1} &=& q_i + p_{i + 1} \qquad \qquad \qquad \mbox{mod}(1)
\end{eqnarray}
The quantized propagator with $N$ discrete levels is given by
\begin{eqnarray}
      \langle n^\prime | \hat{U} | n \rangle &=& {1 \over \sqrt{\rmi N}}
        \exp [\rmi \pi (n - n^\prime)^2 / N] \nonumber \\
      && \qquad \qquad \times \exp \left( \rmi {k N \over 2 \pi} \cos [2 \pi
(n^\prime + a) / N] \right)
\end{eqnarray}
where $n,n^\prime = 0,\dots, N - 1$ and $a$ is a phase term which we set equal
to zero.  The effective Planck constant is $h = 1 / N$ and the average
density of states is $\overline{d} = N / 2 \pi$.

  From the considerations of Sect.~2, examples of nearly pure Gaussian
or exponential decay can be found by selecting an appropriate value of
$\epsilon$ using equation~(\ref{criter}).  In figure~\ref{fig:gaussian},
we see that for small $\hbar$ (large $N$), even without initial state
averaging, the theoretical curves for the Gaussian and exponential regimes
respectively match the quantum results. The theory curves use the
analytic form of the action diffusion constant, which for the standard
map is~\cite{lct}
\begin{equation}
\label{diffu}
     K(E) \approx [1 + 2 J_2(\lambda)] / 4(2 \pi)^4
\end{equation}
Also, it turns out that the expressions for the perturbation matrix
element variance or level slope variance is different for quantized maps
than for continuous dynamical systems since the Floquet eigenangles are
scaled differently than energy eigenvalues.  Here, the variance of the
eigenangle slopes is
\begin{equation}
     \sigma^2_\phi = \sigma^2_v / \hbar^2
     = g (4 \pi)^2 N K(E) \ \ \ \mbox{where} \ g = 2 \ \ \ \beta=1
\end{equation}

In figure~\ref{fig:lorentzian}, we show an example of the behavior of
$f$ as a function of the transition parameter $\Lambda$.  Ergodicity
allows the ensemble averaging to be replaced by an average over
eigenstates.  For quantum maps, the transition parameter is defined in
terms of the variance of the eigenphases as opposed to the eigenenergies
as in continuous systems.   Therefore, for the quantized standard map
\begin{equation}
\label{lambd}
    \Lambda =  {\epsilon^2 \beta \sigma^2_\phi \overline{d}^2 \over 2 g^2}
    = {2 \epsilon^2 N^3 K(E) \over g} \approx
    {\epsilon^2 N^3 \over 8 (2 \pi)^4} [1+2J_2(\lambda)]
\end{equation}
Using the GOE interpolation expression for $f$ in equation~(\ref{fl})
gives excellent agreement with the quantum results.

Finally, we show an example in figure~\ref{fig:crossover} where $\epsilon$
has been chosen so that $A(\epsilon;t)$ exhibits both exponential and
Gaussian behaviors in different time regimes.  The theoretical
expression given in equation~(\ref{eq:full}) with equation~(\ref{diffu})
and the interpolation approximation for $f$ of the previous section
compares well over the full time dependence.  It is possible to isolate
both the exponential and Gaussian components of $A(\epsilon;t)$, and
these are shown in the lower panels. The intercepts of the curves at
$t=0$ show the accuracy of the numerical fit to $f$, and the decay scales
can be seen to be given properly as well.

\section{Conclusions}

The fidelity is extremely useful for studies of decoherence,
reversibility, and tomography in several contexts, and has received a
great deal of attention recently.  It is to be expected that RMT theory
would be helpful in understanding its behavior for strongly chaotic
systems.   Indeed, we applied RMT to develop a uniform approximation that
covers both previously known regimes of Gaussian and exponential decay,
and the theoretical predictions describe the behavior of the quantized
standard map extremely well.  Although not noted earlier in the text, the
range of $\Lambda$ over which the bare degenerate perturbation theory is
accurate, is disappointingly narrow.  However, it helped to provide a
schema for an improved interpolation formula.  We further included the
semiclassical analysis required to show that only one parameter (other
than the density of states/phase space volume) determined the full
behavior, the classical action diffusion constant.  In quantum terms, the
sole information is carried by the transition parameter or alternatively
by the perturbation's spreading width.  As a function of this parameter,
the uniform approximation gives a continuous family of universal decay
curves for all strongly chaotic systems.

Our numerical tests of the time-reversal-invariant uniform approximation
with the quantum kicked rotor relied on large dimensionality examples,
i.e.~$N=1000$, to show that although we averaged over initial states to
generate the theory, the ergodic properties of RMT imply that the results
apply individually to nearly any initial state of a strongly chaotic
system.  For smaller values of $N$, the fluctuations due to not making an
initial state averaging increase and are readily visible.  There
averaging would be required to have the same high quality comparison
between the theoretical prediction and the quantum results.  We would
expect a similar quality of results to apply for
time-reversal-noninvariant chaotic systems.

The new interpolation formulae given in equation~(\ref{fl}) for the
overlap of a perturbed and an unperturbed eigenstate are improvements
over the simple Lorentzian forms since they contain the correct limiting
forms, and are more faithful to the Monte Carlo simulations.
Nevertheless, the previous forms work surprisingly well; even where
the slope at small perturbation goes to infinity in the GOE results,
the previous form never goes too far astray.  The overlap $f$ is important
in other contexts as well.  For example, in mesoscopic physics $f$ is
related to the conductance of a quantum dot undergoing a parametric
change.  For reasons such as this, it would be interesting to know the
full analytic transition with perturbation strength between unity and
zero.

For integrable, near-integrable, and mixed phase space systems, the
behavior of the fidelity is not universal and depends upon the initial
state chosen.  Similarly, chaotic systems in the Lyapunov regime (not
studied in this paper) could also exhibit initial state dependent
phenomena.  Such systems are also of interest and contain a variety of
different regimes; see the work in~\cite{prosen:02,jacquod:02}.  We are
currently pursuing several open questions in these dynamical systems.

\ack

We gratefully acknowledge important discussions with Ph. Jacquod and
C. Lewenkopf, and generous support from ONR Grant N00014-98-1-0079 and NSF
Grant PHY-0098027.

\appendix

\section{Monte Carlo simulations of $f$ for the GOE and GUE}

We use the Hamiltonian
\begin{equation}
   H = \sin(\epsilon) H_0 + \cos(\epsilon) H_1
\end{equation}
in the Monte Carlo simulations.  $H_0$ and $H_1$ are random $(300 \times
300)$  matrices in either the GOE or GUE symmetry.  New random matrices
are  recalculated for each $\epsilon$ to generate independent samples.
An ensemble of 300 such matrices are used where the average density of states,
$\overline{d}$, and variance of the level velocities, $\sigma^2_E$,
are averaged over the entire ensemble.  Only the middle third of the
spectrum is used.  The eigenfunctions of $H_0$
are overlapped with those $H$ and then the absolute value square is taken
to obtain $f$.  $\Lambda$ is given by
\begin{equation}
     \Lambda = \left\{ \matrix{\epsilon^2 \sigma^2_E \overline{d}^2 / 2
     & \qquad \mbox{GOE} \cr \cr
     \epsilon^2 \sigma^2_E \overline{d}^2
     & \qquad \mbox{GUE}}\right.
\end{equation}

\Figures

\begin{figure}
    \caption{\label{fig:interpolation} Overlap of eigenstates, $f$.
    The two upper and two lower plots are for GOE and GUE statistics,
    respectively.  The solid curves in the upper and lower middle plots
    are results from RMT calculations for an ensemble of 300 random
    ($300 \times 300$) matrices.  The dashed curves are the uniform
    interpolation, while the dotted curves are the lorentzian ansatz of
    reference~\cite{alhassid:95}.  The upper middle and lower plots
    show the difference between RMT and the uniform interpolation
    (solid) and the difference between RMT and the lorentzian ansatz
    (dashed).}
\end{figure}

\begin{figure}
    \caption{\label{fig:gaussian}  Example of Gaussian and exponential
    decays.  The quantum standard map curves are solid and the theoretical
    curves are dashed.  The upper plot is for $\epsilon = 2 \times 10^{-3}$ and
    the lower plot is for $\epsilon = 10^{-4}$.  The other parameters are
    $N = 1000$ and $\lambda = 18$.}
\end{figure}

\begin{figure}
    \caption{\label{fig:lorentzian}  Overlap of eigenstates, $f$, for
    the quantum standard map.  The solid curve is for the
    quantum standard map averaged over eigenstates for
    $N = 1000$ and $\lambda = 18$.  The dashed curve is the interpolation
    formula using RMT and degenerate pertubation theory.}
\end{figure}

\begin{figure}
    \caption{\label{fig:crossover}  Crossover decay.  The upper panel
    demonstrates the agreement between the quantum standard map (solid
    curve) and the theoretical results (dashed curve).  The middle and
    lower panel shows each piece of the fidelity (diagonal and off-diagonal)
    separately and the corresponding theoretical results (Gaussian
    and exponential).  The parameters are $N = 1000$, $\lambda = 18$
    and $\epsilon = 5 \times 10^{-4}$.}
\end{figure}


\section*{References}

\begin{thebibliography}{}

\bibitem{btu} Bohigas 0, Tomsovic S, and Ullmo D 1993 {\it Phys. Rep.} {\bf
223} 43

\bibitem{brody} Brody T A, Flores J, French J B, Mello P A, Pandey A
and Wang S S M 1981 {\it Rev. Mod. Phys.} {\bf 53} 385

\bibitem{zas} Zaslavsky G M 1980 {\it Phys. Rep.} {\bf 80} 157

\bibitem{bb} Berry M V and Balazs N L 1979 {\it J. Phys.} {\bf A 12} 625

\bibitem{bbtv}  Berry M V, Balazs N L, Tabor M and Voros A 1979 {\it
Ann. Phys.}  (N. Y.) {\bf 122} 26

\bibitem{th} Tomsovic S and Heller E J 1993 {\it Phys. Rev. E} {\bf 47} 282

\bibitem{jal:01} Jalabert R A and Pastawski H M 2001 {\it Phys. Rev.
Lett.} {\bf 86} 2490

\bibitem{jac:01} Jacquod Ph, Silvestrov P G and Beenakker C W J 2001
{\it Phys. Rev. E} {\bf 64} 055203(R)

\bibitem{cer:02} Cerruti N R and Tomsovic S 2002 {\it Phys. Rev. Lett.}
{\bf 88} 054103

\bibitem{pro:02} Prosen T 2002 {\it Phys. Rev. E} {\bf 65} 036208

\bibitem{cuc:1} Cucchietti F M, Pastawski H M and Wisniacki D A 2002 {\it
Phys. Rev. E} {\bf 65} 045206

\bibitem{cuc:2} Cucchietti F M, Lewenkopf C H, Mucciolo E R, Pastawski H
M and Vallejos R O 2002 {\it Phys. Rev. E} {\bf 65} 046209

\bibitem{ben:02} Benenti G and Casati G 2002 {\it Phys. Rev. E} {\bf 65}
066205

\bibitem{zurek} Zurek W H 1991 {\it Physics Today} {\bf 44} 36; Giulini D
et al 1996 {\it Decoherence and the appearance of a classical world in
quantum theory} (Springer, Berlin)

\bibitem{nielsen:00} Nielsen M A and Chuang I L 2000 {\it Quantum Computation
and Quantum Information} (Cambridge University Press, Cambridge)

\bibitem{pastawski:95} Pastawski H M, Levstein P R and Usaj G 1995
{\it Phys. Rev. Lett.} {\bf 75} 4310

\bibitem{flatte:79} Flatt\'{e} S M, Dashen R, Munk W H and Zachariasen F 1979
{\it Sound Transmission through a Fluctuating Ocean} (Cambridge University
Press, Cambridge)

\bibitem{altshuler:85} Al'tshuler B L and Spivak B Z 1985
{\it Pis'ma Zh. Eksp. Teor. Fiz.} {\bf 42}, 294 [1985 {\it JETP Lett.}
{\bf 42}, 447]; Feng S, Lee P A and Stone A D 1986 {\it Phys. Rev. Lett.}
{\bf 56}, 1960

\bibitem{meso} Lee P A, Stone A D and Fukuyama H 1987 {\it Phys. Rev. B}
{\bf 35} 1039; Folk J A, Patel S R, Godijn S F, Huibers A G,
Cronenwett S M, Marcus C M, Campman K and Gossard A C 1996
{\it Phys. Rev. Lett.} {\bf 76} 1699

\bibitem{peres:84} Peres A 1984 {\it Phys. Rev. A} {\bf 30} 1610

\bibitem{usaj:98} Usaj G, Pastawski H M and Levstein P R 1998 {\it Mol. Phys.}
{\bf 95} 1229

\bibitem{prosen:02} Prosen T and Znidaric M 2002 {\it J. Phys. A: Math.
Gen.} {\bf 35} 1455

\bibitem{jacquod:02} Jacquod Ph, Adagideli I and Beenakker C W J 2002
{\it Preprint} nlin.CD/0206160

\bibitem{cerruti:03} Cerruti N R and Tomsovic S, unpublished

\bibitem{old} Zhang S, Meier B H and Ernst R R 1992 {\it Phys. Rev. Lett.}
{\bf 69} 2149

\bibitem{fkpt} French J B, Kota V K B, Pandey A and Tomsovic S 1988
{\it Ann. Phys. (N.Y.)} {\bf 181} 198

\bibitem{bgs} Bohigas O, Giannoni M-J, and Schmit C 1984 {\it Phys. Rev.
Lett.} {\bf 52} 1;  1984 {\it J. Physique Lett.} {\bf 45} 1015

\bibitem{pandey} Pandey A 1979 {\it Ann.~Phys.~(N.Y)} {\bf 119} 170; 1978
Doctoral dissertation University of Rochester

\bibitem{feyn} Hellmann H 1937 {\it Einfuhrung in die Quantenchemie}
(Deuieke, Leipzig); Feynman R P 1939 {\it Phys. Rev.} {\bf 56} 340

\bibitem{lct} Lakshminarayan A, Cerruti N R and Tomsovic S
1999 {\it Phys. Rev. E} {\bf 60} 3992; 2000 {\it Phys. Rev. E} {\bf 61}
6031(E)

\bibitem{cllt} Cerruti N R, Lakshminarayan A,  Lefebvre J H and Tomsovic S
2001 {\it Phys. Rev. E} {\bf 63} 016208

\bibitem{keating} Berry M V and Keating J P 1994 {\it J. Phys. A: Math. Gen.}
{\bf 27} 6167

\bibitem{bgos} Bohigas O, Giannoni M J, Ozorio de Almeida A M and
Schmit C 1995 {\it Nonlinearity} {\bf 8} 203

\bibitem{bruus:96} Bruus H, Lewenkopf C H and Mucciolo E R 1996
{\it Phys. Rev. B} {\bf 53} 9968

\bibitem{alhassid:95} Alhassid Y and Attias H 1995 {\it Phys. Rev. Lett.}
{\bf 74} 4635; Attias H and Alhassid Y 1995 {\it Phys. Rev. E} {\bf 52}
4776

\bibitem{kusnezov:96} Kusnezov D and Mitchell D 1996 {\it Phys. Rev. C}
{\bf 54} 147

\bibitem{sth} O'Connor P W, Tomsovic S and Heller E J 1992 {\it Physica}
{\bf D 55} 340; Sep\'ulveda M A, Tomsovic S and Heller E J 1992
{\it Phys.~Rev.~Lett.} {\bf 69} 402

\bibitem{as} Simons B D, Lee P A and Al'tshuer B L 1993 {\it Phys.
Rev. Lett.} {\bf 70} 4122

\bibitem{tomsovic:86} Tomsovic S 1986 Univerisity of Rochester Ph.D.
Thesis; 1986 UR Report 974

\bibitem{tomsovic:94} Tomsovic S and Ullmo D 1994 {\it Phys. Rev. E} {\bf
50} 145

\bibitem{wilkinson:95} Wilkinson M and Walker P N 1995 {\it J. Phys. A:
Math. Gen.} {\bf 28} 6143

\end{thebibliography}
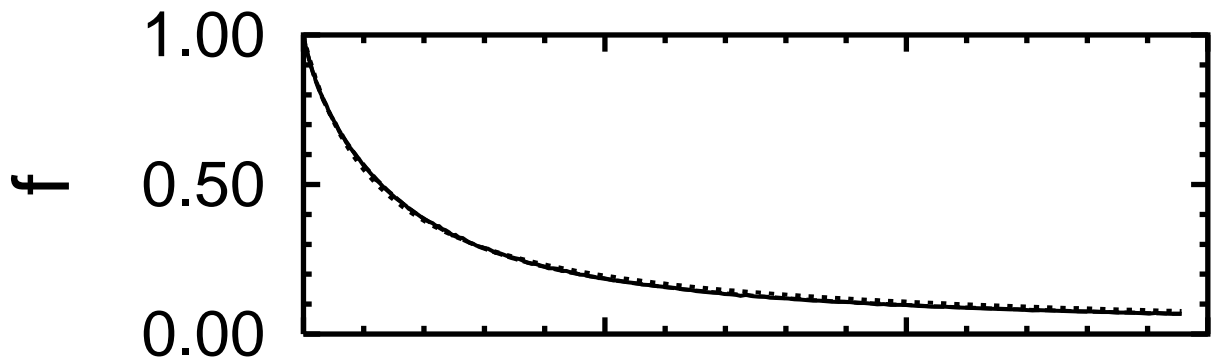
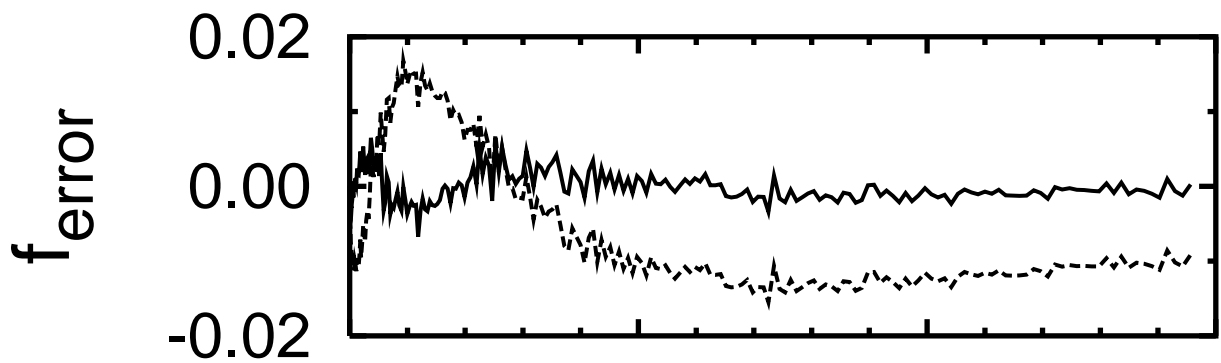
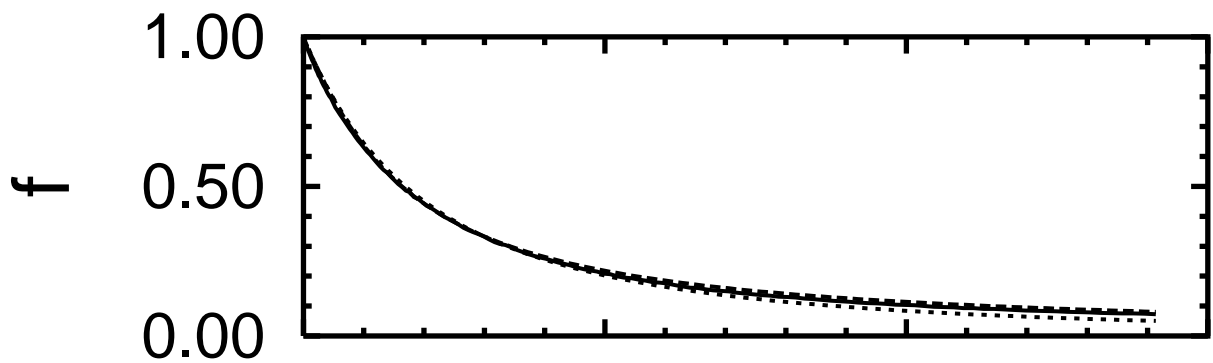
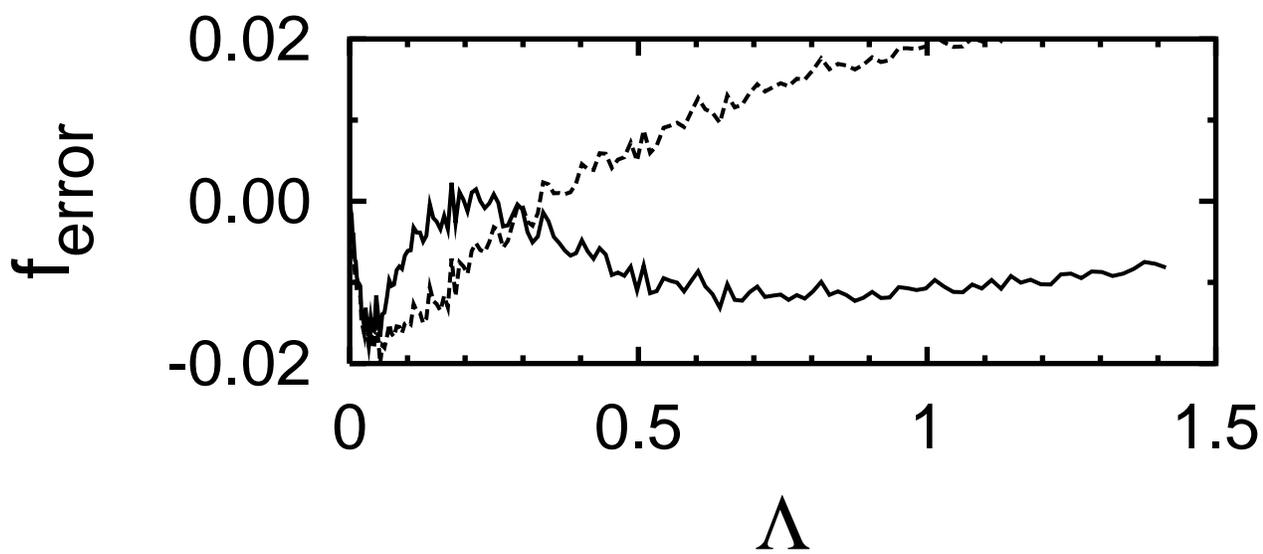
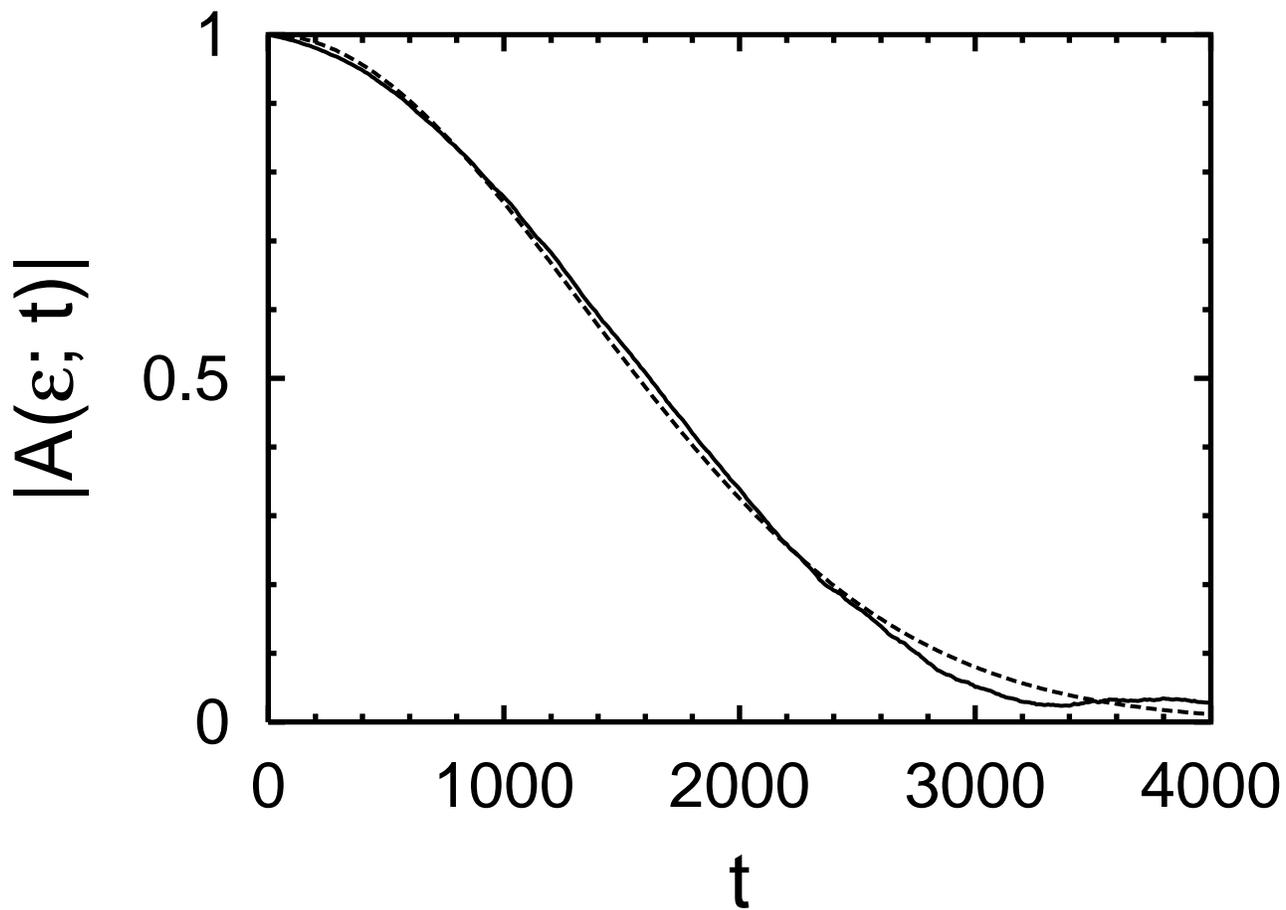
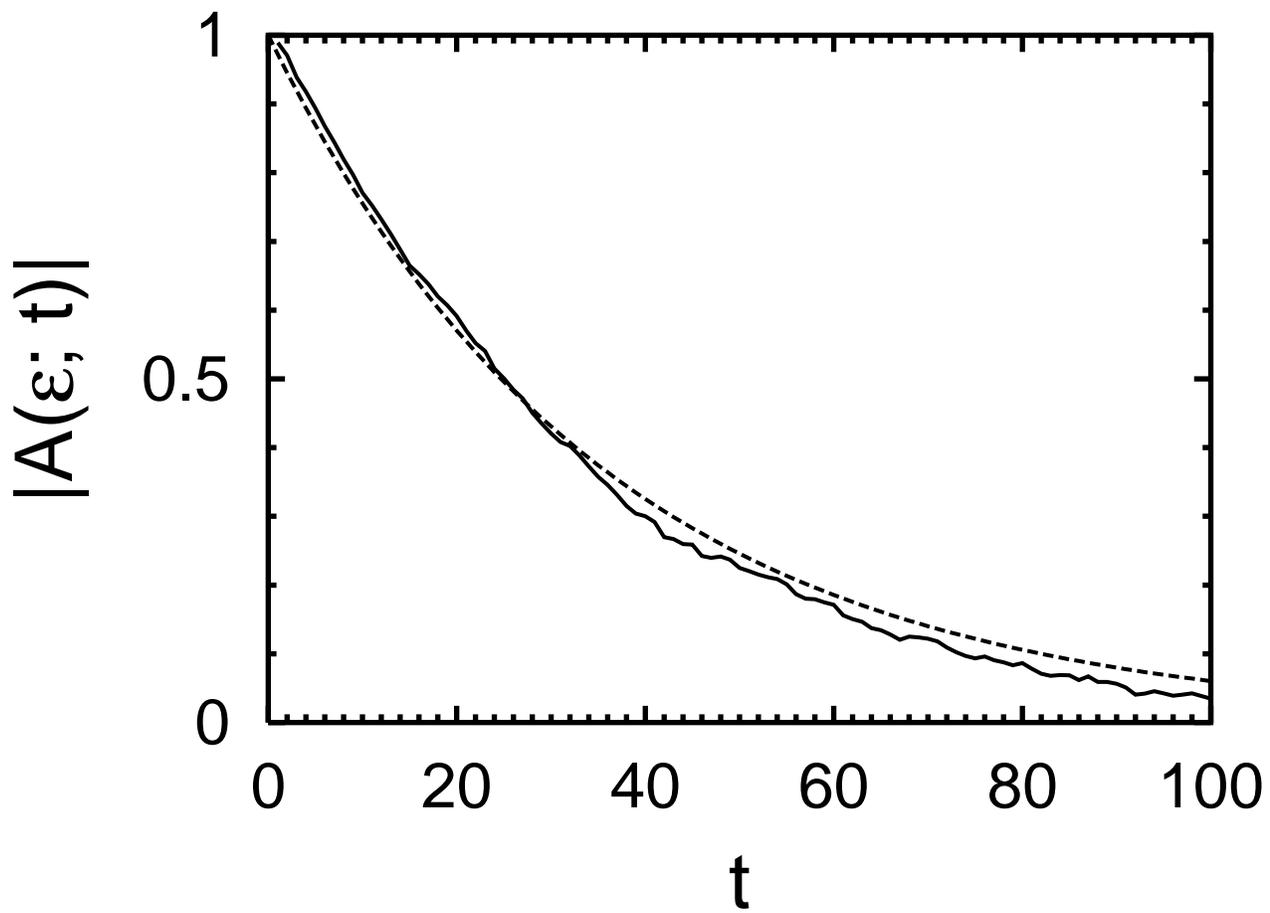
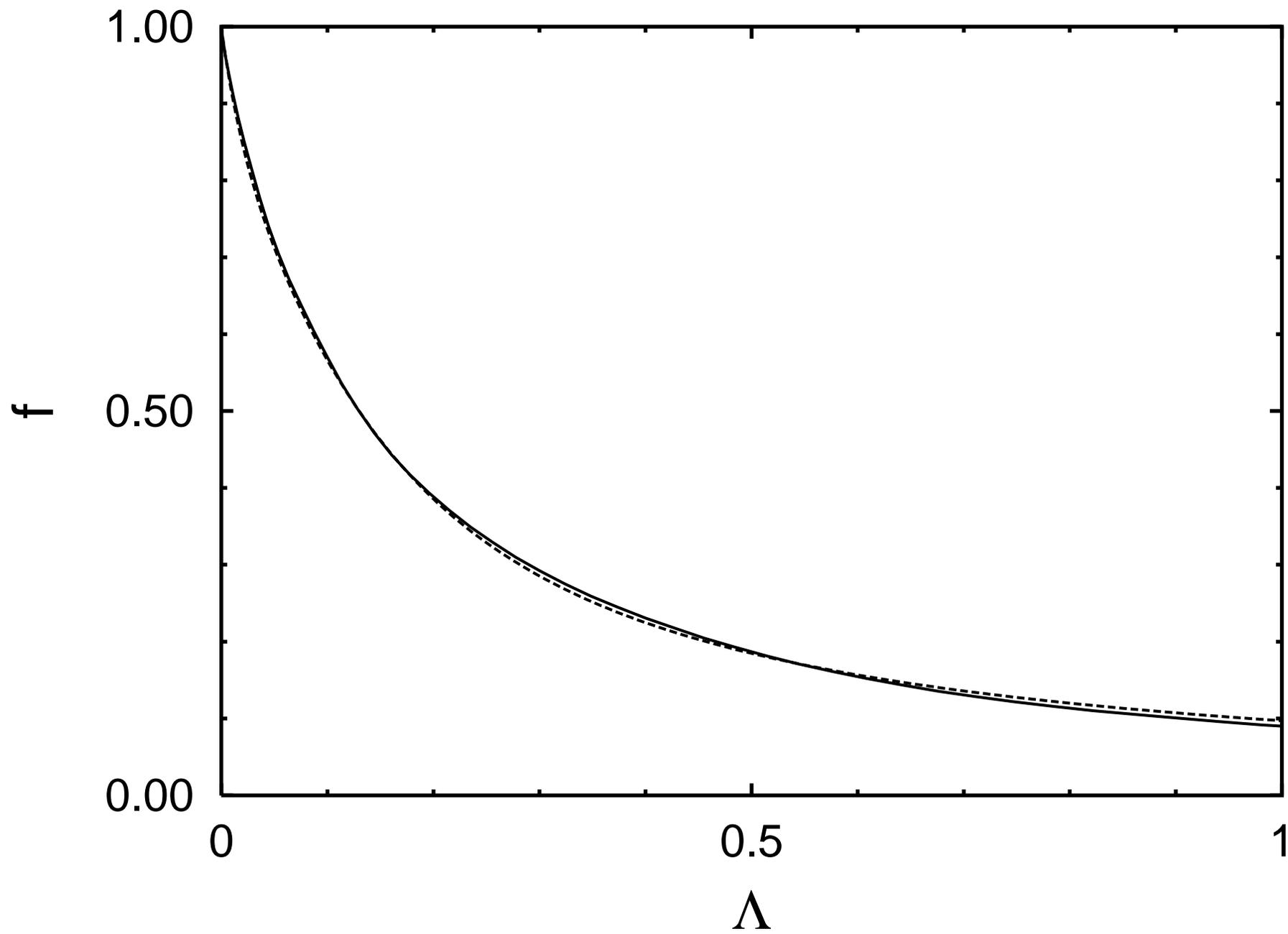
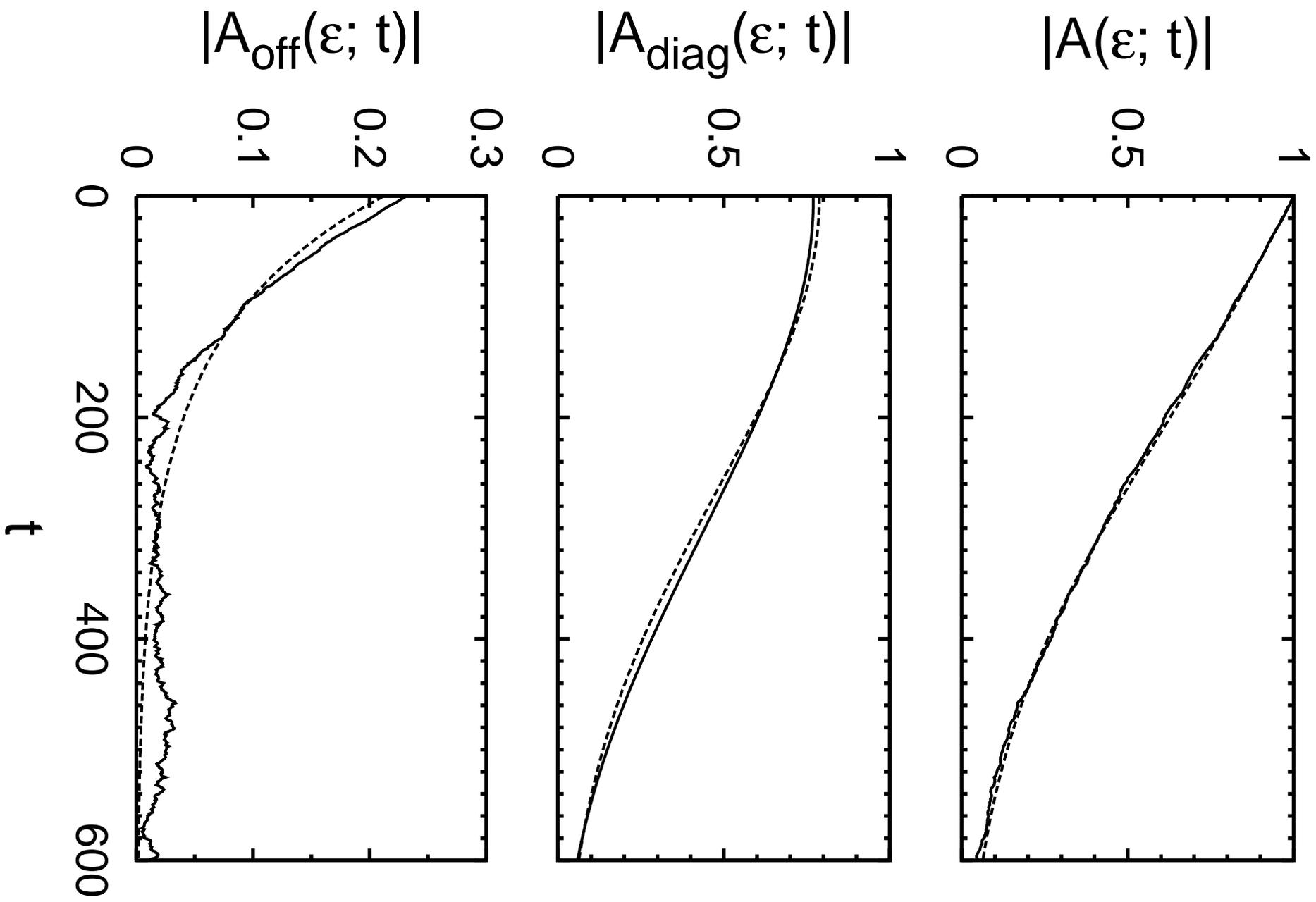
\end{document}